\RequirePackage{lineno}
 \documentclass{ws-ijmpe}
 \usepackage{lineno}
\usepackage[super,compress]{cite}
\usepackage{graphicx}
\modulolinenumbers[1]
\begin{document}
\markboth{Chunjian Zhang}{An overview of new measurements of flow, chirality, and vorticity from STAR experiment}

%
\catchline{}{}{}{}{}
%

\title{An overview of new measurements of flow, chirality, and vorticity from STAR experiment
}

\author{Chunjian Zhang\footnote{
Speaker (10th International Conference on New Frontiers in Physics (ICNFP 2021), Kolymbari, Crete, Greece)
} \space (for the STAR Collaboration)
}

\address{Chemistry Department, Stony Brook University, Stony Brook,\\
New York, 11794, USA \\
chun-jian.zhang@stonybrook.edu}

\maketitle


\begin{abstract}
In relativistic heavy-ion collisions, the properties of quark-gluon plasma (QGP) and complex dynamics of multi-scale processes in Quantum Chromodynamics (QCD) are studied by  analyzing the final state produced particles in a variety of different ways. In these proceedings, we present an overview of new detailed measurements of flow, chirality, and vorticity by the STAR experiment at RHIC.  Furthermore, STAR’s future opportunities for the precision measurements on small systems, fixed-target (FXT) mode, and Beam Energy Scan (BES-II) program are discussed.
\end{abstract}

\ccode{PACS numbers: 26.75.Nq, 25.75.Ld, 25.75.Ag, 25.75.G}


\section{Introduction}	
The initial conditions and dynamics of a hot and dense phase of QCD matter, the strongly interacting QGP \cite{Romatschke,Teaney}, is naturally created in the nuclear collisions at the Relativistic Heavy Ion Collider (RHIC) and the Large Hadron Collider (LHC). After the collision, the subsequent fluid motion and the expansion of QGP flow hydrodynamically \cite{Heinz,FGG,HuichaoPRL}, later on the QGP turns into a lower-temperature hadronic phase. Thus, such nuclear collisions offer an ideal environment to explore fundamental physics. During the QGP fireball expansion, spatial anisotropies in the initial state, lead to final state momentum anisotropies. The large azimuthal modulations in the final distributions of the produced particles, known as collective flow phenomena, are typically characterized by Fourier coefficients \cite{SV}.

It is also of fundamental importance to explore and understand the topological and electromagnetic properties of QGP. In the early stage of the nuclear collisions, a strong electromagnetic field exists and could induce an electric current along the direction of the strong magnetic field $\mathbf{B}$ for chirality imbalanced domains with a nonzero topological charge inside the hot chiral-symmetric QGP, which is known as Chiral Magnetic Effect (CME) \cite{DK,JF,SAV}. The search for the CME, in such a unique micro-universe environment created by relativistic nuclear collision experiments, has been pursued for more than a decade. 

The non-central heavy-ion collisions have large orbital angular momentum that could result in strong fluid shear and nonvanishing local fluid vorticity \cite{Liang,FB}. In such vorticity of the fluid cell, the spin-orbit coupling effect could lead to preferential orientation of particle spins along the direction of local fluid vorticity \cite{FB2,Huichao2,Pang}. The first measurement of final state $\Lambda$ hyperon polarization by STAR \cite{Nature} sheds light on such vortical structure and its transport properties. Measurements of $\Xi$ and $\Omega$ hyperons polarizations \cite{PRLTaku}, $\Lambda$-hyperon polarization at lower BES energies and FXT collisions\cite{Joey}, and the theoretical model calculations \cite{FB2,Huichao2,Pang,Huang,Sun,Deng} are crucial for understanding the vorticity and polarization phenomena.

In these proceedings, we present recent measurements of the flow, chirality, and vorticity measurements by the STAR experiment at RHIC, and discuss future opportunities. 

\section{Flow and Fluctuations }
\subsection{Anisotropic Flow in Small Systems}
The origin of a sizeable azimuthal anisotropy in small systems is still unknown, although the anisotropic flow for different harmonics and different particle species have been extensively measured via two- and multi-particle correlations at RHIC \cite{star0,phenix0,phenix1} and the LHC \cite{CMS,ATLAS,ALICE}. 
Some of the unsolved questions in understanding the behavior of small system collisions are 1) what determines the initial geometry? 2) what is the connection between initial state and final state correlations? 3) what are the roles of nucleonic and sub-nucleonic fluctuations?
 
\begin{figure}[hbpt]
\centerline{\includegraphics[width=0.99\textwidth]{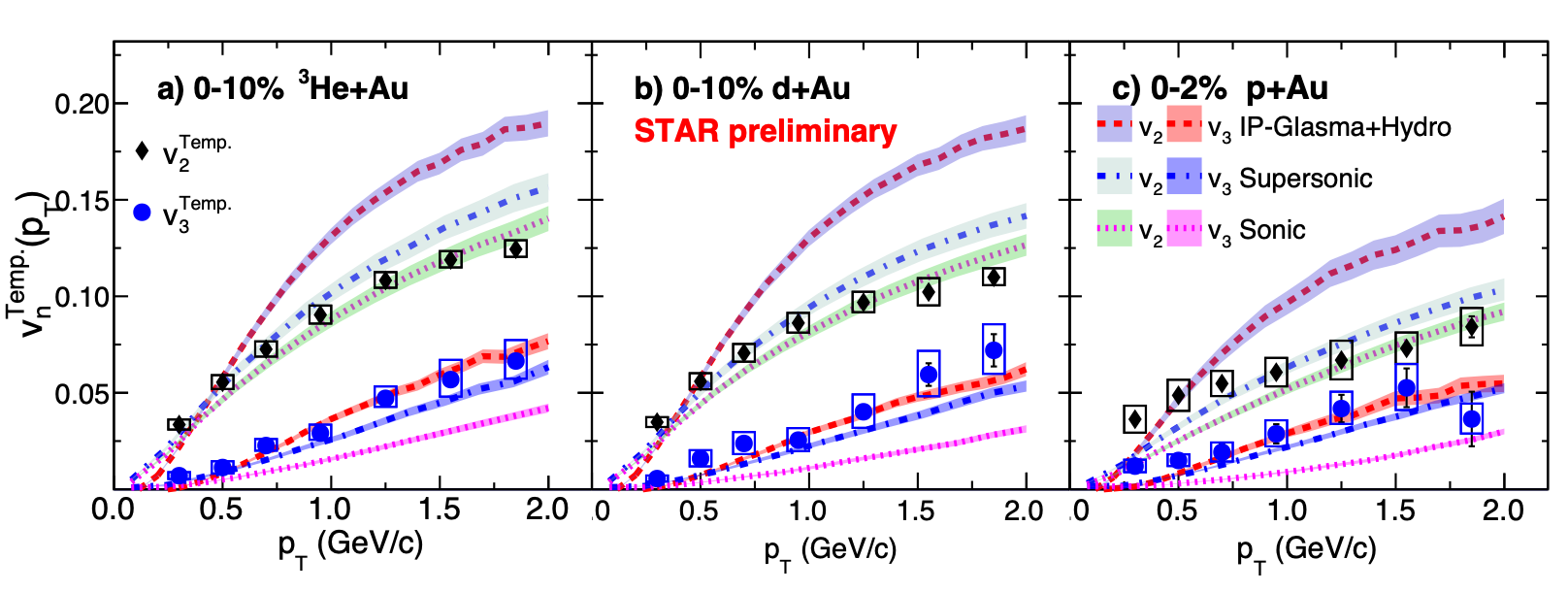}}
\caption{Comparison of $v_{2,3}$($p_T$) values in the central $p/d/^3$He+Au collisions at $\sqrt{s_{NN}}$ = 200 GeV with the calculations from Sonic \cite{sonic}, Supersonic \cite{supersonic}, and IP-Glasma+MUSIC+UrQMD \cite{ipg} calculations.
	\label{f1}}
\end{figure}
Figure \ref{f1} shows $v_{2,3}$($p_T$) from template-fit method \cite{temp} and the comparisons with Sonic \cite{sonic}, Supersonic \cite{supersonic}, and IP-Glasma+MUSIC+UrQMD \cite{ipg} calculations. The measurements from STAR Collaboration show a hierarchy: $v_2^{p+\mathrm{Au}} < v_2^{d+\mathrm{Au}} \sim v_2^{^3\mathrm{He}+\mathrm{Au}}$ and $v_3^{p+\mathrm{Au}} \sim v_3^{d+\mathrm{Au}} \sim v_3^{^3\mathrm{He}+\mathrm{Au}}$. The Sonic calculations with initial geometry eccentricity from nucleon Glauber model predict the $v_2$ well but underpredict $v_3$ in $p/d/^3$He+Au collisions. After including the pre-equilibrium flow, the Supersonic calculations match the $v_n$ better. The calculations from the IP-Glasma+MUSIC+UrQMD model with sub-nucleonic fluctuations over-predict the $v_2$, while it reproduces the $v_3$ in the three collision systems.
\begin{figure}[hbpt]
\centerline{\includegraphics[width=0.85\textwidth]{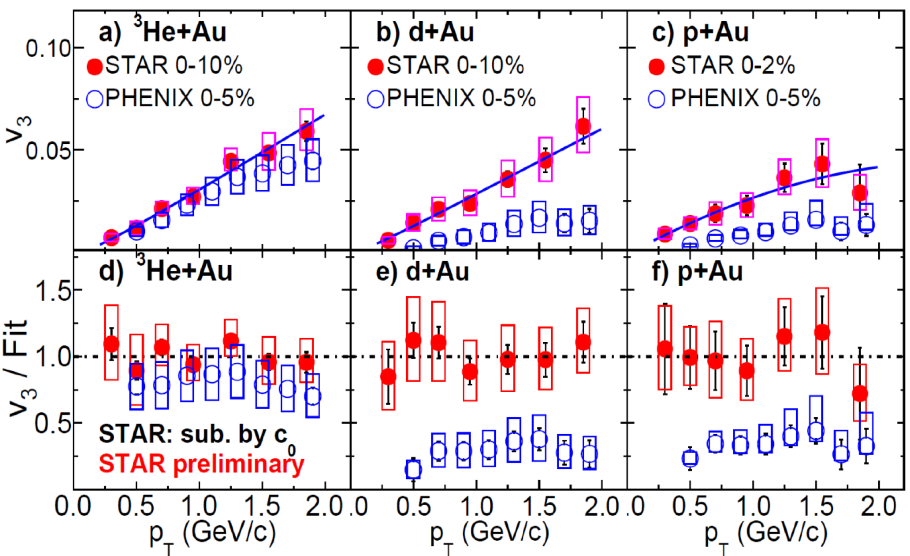}}
\caption{Comparison of $v_{3}$($p_T$) measurements obtained by STAR and PHENIX in the central $p/d/^3$He+Au collisions at $\sqrt{s_{NN}}$ = 200 GeV. The solid lines in the top panels represent a fit to the STAR data. The bottom panels show the ratio of the respective data to this fit.  \label{f2}}
\end{figure}

Figure \ref{f2} presents the measurements of $v_{3}$($p_T$) from peripheral subtraction method \cite{roy} in the central $p/d/^3$He+Au collisions at $\sqrt{s_{NN}}$ = 200 GeV at STAR, and are compared to published PHENIX measurements \cite{PHENIX}. $v_{3}$($p_T$) shows a reasonable agreement in $^3$He+Au collisions between STAR and PHENIX. However, within the statistical and systematic uncertainties, there is a factor 3 - 4 discrepancy in $p/d$+Au collisions between STAR and PHENIX measurements. STAR results imply, the fluctuation-driven $v_{3}$($p_T$) is system-independent. The final result has been reported in Ref. \cite{STARsmall}. Future measurements including proper nonflow treatment, enhanced detector acceptance, and various other collisions, such as O+O, could provide additional constraints and insights on the origin of QGP fluidity in small system collisions. 

\subsection{Flow Correlations with Mean Transverse Momentum}
The correlation between flow harmonics ($v_n$) and the mean transverse momentum ($[p_T]$), estimated by using a Pearson correlation coefficient $\rho(v_n^2,[p_T])$, is proposed to reveal interesting information both on the correlations in the initial state between the geometric size and the eccentricities \cite{Piotr}. In relativistic heavy-ion collisions, the shape and size of the QGP may depend on the fluctuations and the shape of the colliding nuclei where the spatial distribution of nucleons is often described by a Woods-Saxon density profile: $\rho(r, \theta)=\frac{\rho_{0}}{1+e^{\left[r-R_{0}\left(1+\beta_{2} Y_{2,0}(\theta)+\beta_{3} Y_{3,0}(\theta)\right) / a_{0}\right]}}$, where $\rho_0$ denotes the nucleon density at the center of the nucleus, $R_0=1.2A^{1/3}$ is the nuclear radius, and $a_0$ is the surface diffuseness parameter (known as skin depth). $\mathrm{Y_{n,0}}(\theta)$ (n=2,3) are spherical harmonics. Therefore, $\rho(v_n^2,[p_T])$ is of particular interest is to distinguish the information of the initial geometry effect induced by the nuclear deformation \cite{GG1,JJ1, GG5, GG6}.

In the hydrodynamic calculations, elliptic flow ($v_2$) emerges as a response to the initial eccentricity with $v_2 = k_2\epsilon_2$. This leads to an enhanced fluctuations of the observed $v_2$ \cite{Shou,GG2,GG3} in collisions of deformed nuclei. As shown in the talk \cite{ICNFPtalk}, the expected anticorrelation between $v_2$ and normalized average transverse momentum $\left\langle p_{T}\right\rangle/\left\langle \left\langle p_{T}\right\rangle \right\rangle-1$ is observed in collisions of prolate $^{238}$U nuclei. On the other hand, in collisions of oblate $^{197}$Au, $v_2$ is observed to be essentially flat with only a slight increase of $v_2$ with $\left\langle p_{T}\right\rangle/\left\langle \left\langle p_{T}\right\rangle \right\rangle-1$ due to the increasing impact parameter. Note that TRENTo with initial state fluctuations can capture the trend for Au+Au collisions.

The sign-change of Pearson correlation coefficient $\rho(v_2^2,[p_T])$ in U+U collisions is observed towards central collisions, whereas the result from Au+Au collisions is positive throughout. $\rho(v_3^2,[p_T])$, which is expected to be fluctuation driven, is almost identical between Au+Au and U+U collision systems across the whole centrality range. The comparison of $\rho(v_2^2,[p_T])$ with state-of-the-art hydrodynamic calculations shows hierarchical trends, and suggests the most striking signature of nuclear deformation $\beta_2$ of $^{238}$U to be around 0.3, observed for the first time in high-energy nuclear experiments so far. Moreover, to further constrain the initial conditions and transport properities in hydrodynamic evolution, the experiments at the LHC \cite{vnpt1,vnpt2,vnpt3} and phenomenological studies \cite{GG1,BS,vnpt4,vnpt5} have also reported the studies of the $\rho(v_n^2,[p_T])$. In future, such calculations also could be conducted in the Ru+Ru and Zr+Zr collisions, since the final state effects are totally canceled \cite{ratiosI}. 

\subsection{Transverse Momentum Fluctuations}
In relativistic heavy-ion collisions, the event-by-event mean transverse momentum fluctuations are sensitive to overlap area and energy density fluctuations in the initial state \cite{GGpt}. Therefore, the shape of the nucleus could also be imaged and the size fluctuations could be used to isolate the $\beta_2$ dependence, especially in central and ultra-central collisions. The analytical estimation of shape and size are strongly correlated with nuclear deformation as illustrated in Ref. \cite{JJ2}. So far, experimental measurements are limited to the mean transverse momentum and and its variance \cite{pT1,pT2,pT3,pT4,pT5}. To save the computational overhead from loop-calculations, a framework for calculating the higher-order dynamical $p_T$ cumulants up to fourth-order using the standard and subevent methods is established and detailed in Ref. \cite{Som}.

As shown in the talk \cite{ICNFPtalk}, the normalized variance $\sqrt{\left\langle\left(\delta p_T\right)^2\right\rangle} /\left\langle p_T\right\rangle$ approximately follows a power-law dependence as a function of multiplicity owing to dynamical correlations on top of correlations arising from independent superposition picture \cite{Wang}. The additional fluctuation induced by the nuclear deformation $\beta_2$ of $^{238}$U collisions is observed as an enhancement in normalized variance $\sqrt{\left\langle\left(\delta p_T\right)^2\right\rangle} /\left\langle p_T\right\rangle$, normalized skewness $\left\langle\left(\delta p_T\right)^3\right\rangle\left\langle\left\langle\delta p_T\right)^2\right\rangle^{3/2}$, and intensive skewness $\left\langle\left(\delta p_T\right)^3\right\rangle^*\left\langle p_T\right\rangle /\left\langle\left(\delta p_T\right)^2\right\rangle^2$. Interestingly, the normalized kurtosis $\left\langle\left(\delta p_T\right)^4\right\rangle_c /\left\langle\left(\delta p_T\right)^2\right\rangle^2$ in U+U collisions shows a clear and significant sign-change behavior in ultracentral regions. Remarkably, $p_T$ fluctuations from mean to kurtosis could be used as a complementary tool to probe nuclear structure in $^{238}$U, $^{96}$Ru, and $^{96}$Zr with heavy-ion colliders in the future \cite{JJ2}.

\subsection{Longitudinal Flow Decorrelations}
Initial state fluctuations and final state dynamics of QGP are important properties in heavy-ion collisions. The distributions of particle production sources and the associated eccentricity, fluctuate along the pseudorapidity ($\eta$), which causes a non boost-invariant flow, known as flow decorrelations. \cite{decorr1,decorr2,decorr3,decorr4}. The flow decorrelations are usually quantified by the factorization ratio $r_{n}(\eta)=\frac{\left\langle\mathbf{q}_{n}(-\eta) \mathbf{q}_{n}^{*}\left(\eta_{\mathrm{ref}}\right)+\mathbf{q}_{n}(\eta) \mathbf{q}_{n}^{*}\left(-\eta_{\mathrm{ref}}\right)\right\rangle}{\left\langle\mathbf{q}_{n}(\eta) \mathbf{q}_{n}^{*}\left(\eta_{\mathrm{ref}}\right)+\mathbf{q}_{n}(-\eta) \mathbf{q}_{n}^{*}\left(-\eta_{\mathrm{ref}}\right)\right\rangle}$ where the flow vector $\mathbf{q}_{n} \equiv \sum \omega_{i} e^{i n \phi_{i}} / \sum \omega_{i}$ and $\omega_{i}$ is the efficiency correction \cite{decorr5,decorr6,decorr7}. To improve the understanding of the longitudinal structure, a broad range of energy dependence of longitudinal flow decorrelations from the LHC to RHIC is crucial.

\begin{figure}[hbpt]
\centerline{\includegraphics[width=0.6\textwidth]{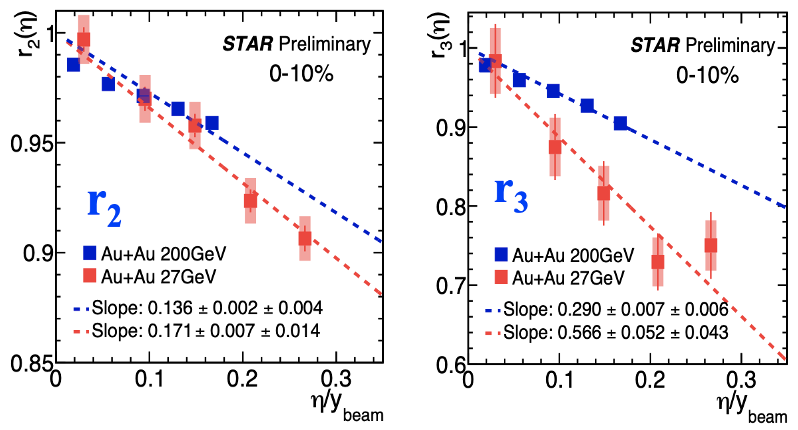}}
\caption{The $r_2(\eta)$ (left panel) and $r_3(\eta)$ (right panel) as a function of $\eta/y_{\mathrm{beam}}$ in 0-10\% in Au+Au collisions at 27 and 200 GeV. A linear fit is in dashed line. \label{f6}}
\end{figure}
In STAR, such an analysis was performed using the charged particles with $0.4<p_T<4$ GeV/c from the Time Projection Chamber (TPC, $|\eta|<1$), and the reference flow vector is calculated from the Event Plane Detector (EPD, 2.1 $<|\eta_{\mathrm{ref}}|< 5.1$) and Forward Meson Sepctrometer (FMS, 2.5 $<|\eta_{\mathrm{ref}}|< 4$) for $\sqrt{s_{NN}}=$ 27 and 200 GeV Au+Au collisions, respectively. To investigate the energy dependence of flow decorrelations, a comparison between 27 and 200 GeV has been shown in Fig. \ref{f6} with a beam rapidity normalization. The $r_2$ shows slight energy dependence while $r_3$ shows clear energy dependence and a stronger decorrelation at 27 GeV after beam-rapidity normalization. In future, collision energy scan using high statistics BES-II data and system-size scan would help to better understand the logitudinal dynamics in heavy-ion collisions.

\subsection{Collectivity measurements at FXT Program}
To study the possible first-order phase transition and a QCD critical point, the BES-I and BES-II data-taking \cite{bes2,besPRL} with adequate luminosity were achieved. Moreover, RHIC also pursued the FXT heavy-ion program \cite{fxt1,fxt4} at high baryon density region, by inserting a gold target into the beam pipe and circulating one beam, to broaden the reach of BES-II data-taking and allow the STAR to access energies below $\sqrt{s_{NN}}$=7.7 GeV.

\begin{figure}[hbpt]
\centerline{\includegraphics[width=0.99\textwidth]{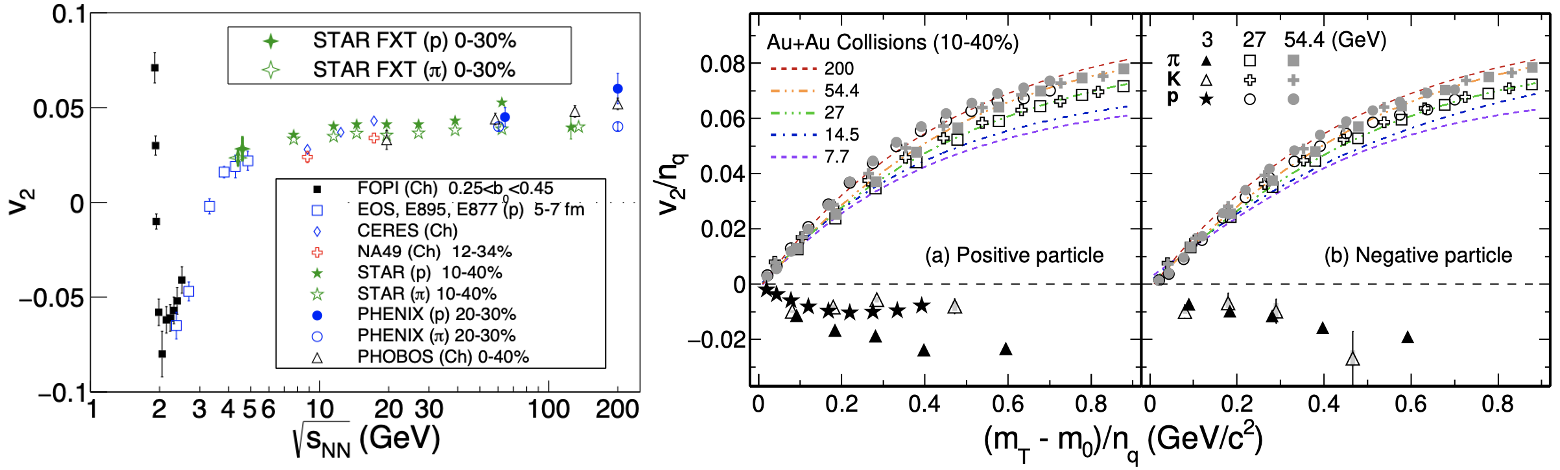}}
\caption{$v_2$ measured by several experiments and STAR 4.5 GeV Au+Au FXT points for protons and pions are near the transition region \cite{fxt2}. $v_2$ scaled by the number of constituent quarks ($v_2/n_q$) as a function of the scaled transverse kinetic energy ($\left(m_{T}-m_{0}\right) / n_{q}$) for pions, kaons and protons from 3 GeV Au+Au FXT \cite{fxt3}. \label{f7}}
\end{figure}

Left panel in Fig. \ref{f7} shows the measurements of the beam energy dependence of elliptic flow $v_2$ for all charged particles integrated over $p_T$. The current results from 4.5 GeV Au+Au FXT are consistent with the trends established by the previously published data for various experiments. From squeeze-out to in-plane elliptic expansion, the $v_2$ changes sign around 3 - 4 GeV collision energies. Such phenomenon has been observed in 3 GeV Au+Au FXT results where $\pi$, $K$ and $p$ are shown by the filled triangles, open triangles, and filled stars in the middle and right panels of Fig. \ref{f7}. The breakdown of number of constituent quark (NCQ) scaling indicates the disappearance of partonic collectivity in such low energy collisions. The detailed model comparisons in Ref. \cite{fxt3} show that partonic interaction is no longer dominant and baryonic scatterings take over at 3 GeV.

\begin{figure}[hbpt]
\centerline{\includegraphics[width=0.95\textwidth]{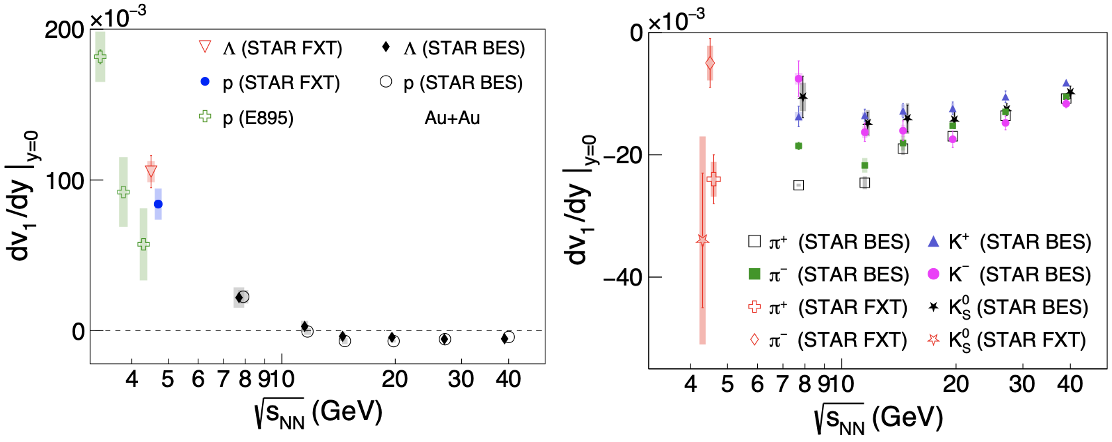}}
\caption{The directed flow slope of $dv_1/dy|_{y=0}$ at midrapidity for baryons (left panel) and mesons (right panel) are measured at 4.5 GeV Au+Au FXT comparing the STAR BES energies and AGS E895 experimental results.\cite{fxt3}. \label{f8}}
\end{figure}

The directed flow reflects the early time expansion, Equation of State (EOS), and the nature of phase transition. Figure \ref{f8} reports the slope of $dv_1/dy|_{y=0}$ for baryons (left) and mesons (right) in 4.5 GeV Au+Au FXT mode. Currently, the results of proton and $\Lambda$ directed flow are in agreement within the uncertainties. The proton $v_1$ agrees with the E895 4.3 GeV energy data within errors. Interestingly, the observed difference between $\pi^{+}$ and $\pi^{-}$ might be due to the isospin effect at lower energy or Coulomb dynamics \cite{fxt2}. 

\section{Chiral Magnetic Effect}
Relativistic heavy-ion collisions can create the strongest electromagnetic field of $eB \sim 10^{18}$ G in the universe \cite{DK1}. An imbalance between the numbers of left- and right-handed (anti)quarks occurs due to the locally violated parity $(\mathcal{P})$ and charge-parity $(\mathcal{C} P)$ symmetries in such a strong $\mathbf{B}$ field \cite{DK2,DK3}. A charge separation along the direction of the magnetic field, a novel CME phenomenon, has been extensively studied using STAR data. \cite{CMEstar0, CMEstar1, CMEstar2}. 

\begin{figure}[hbpt]
\centerline{\includegraphics[width=0.99\textwidth]{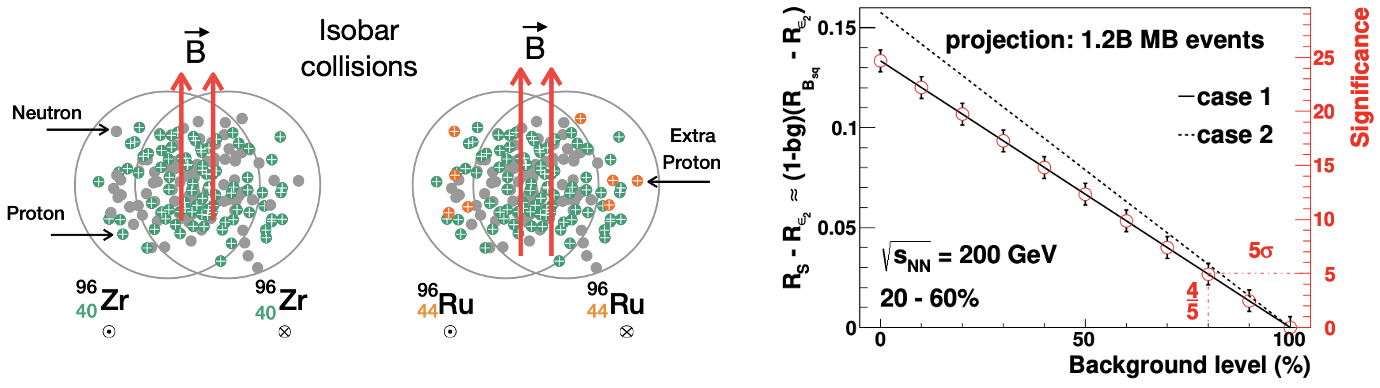}}
\caption{The isobar collisions at RHIC: the stronger magnetic field of Ru+Ru collisions resulting in greater separation of charged particles is expected than Zr+Zr collisions. Magnitude and significance of the relative difference in the projected $\gamma$ correlator between Ru+Ru and Zr+Zr at 200 GeV \cite{CMEstar1}.\label{f9}}
\end{figure}

Figure \ref{f9} left panel shows the cartoon for isobar collisions at RHIC: the stronger magnetic field in Ru+Ru collisions will result in a greater separation of charged particles in Zr+Zr collisions assuming the same background effects in these isobars. The most widely used observable in the CME search is the $\gamma$ correlator, $\gamma=\left\langle\cos \left(\phi_\alpha+\phi_\beta-2 \Psi_{\mathrm{RP}}\right)\right\rangle$, where $\phi_\alpha$ and $\phi_\beta$ are the azimuthal angles of particles of interest (POIs) and $\Psi_{\mathrm{RP}}$ is the reaction plane. The magnitude (left axis) and significance (right axis) of the projected difference in $\gamma$ correlator in isobar runs change accordingly as shown in Fig. \ref{f9} right panel when a different background level is assumed. It has been proposed to be able to determine the CME signal with 5 $\sigma$ significance if 1.2 billion events for each collision system at 200 GeV are taken. The details of the observables for CME search can be found in Ref. \cite{CMEstar} no CME signature that satisfies the predefined criteria has been observed in this blind isobar analysis by STAR. However, the future unblinded analysis with more comprehensive baselines, background estimations from the difference of nuclear structure, and further endeavors based on BES-II data are still ongoing.

\begin{figure}[hbpt]
\centerline{\includegraphics[width=0.9\textwidth]{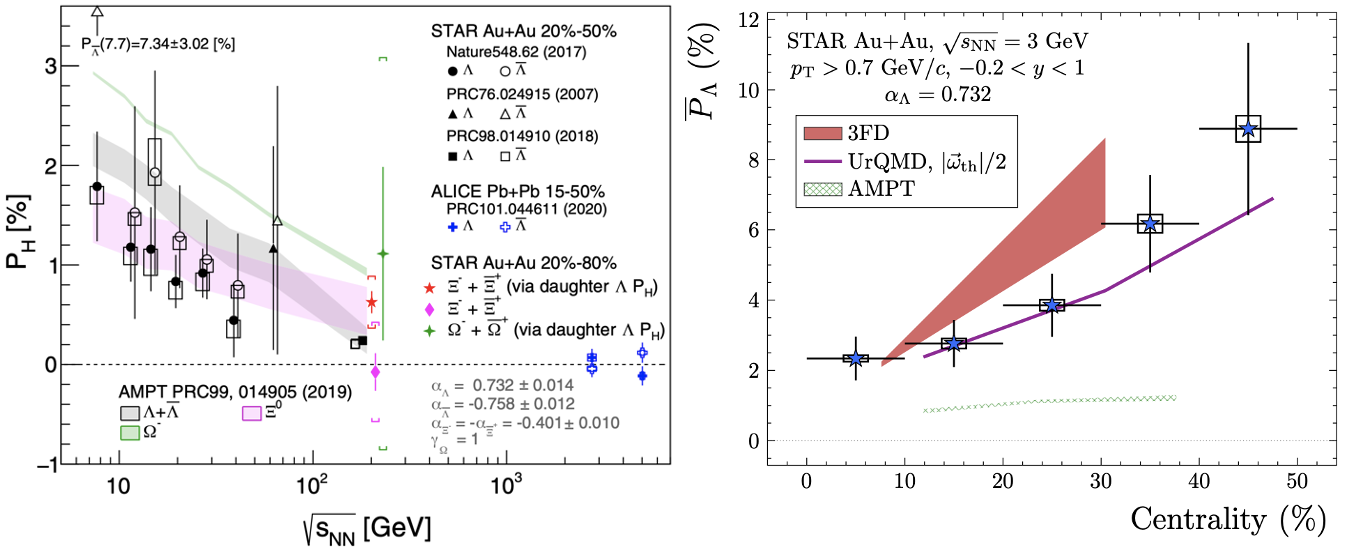}}
\caption{The energy dependence of the hyperon global polarization measurements with the newly added $\Xi$ and $\Omega$ in 200 GeV Au+Au 200 results in the left panel from Ref. \cite{PRLTaku}. The centrality dependence of $\bar{P}_{\Lambda}$ in Au+Au 3 GeV FXT mode comparing to the model calculations from Ref. \cite{Joey} is shown in the right panel. \label{f10}}
\end{figure}

\section{Vorticity and Polarization}
Experimental measurements of the hyperon polarization and the theoretical calculations from hydrodynamics and transport models reveal that the QGP is a vortical fluid \cite{Nature,PRLTaku,Joey,Liang,FB,FB2,Huichao2,Pang,Huang,Sun,Deng}. However, many questions are raised including the sign problems in differential measurements of local polarizations, uniform rapidity dependence but energy dependence in global polarization. Whether a significant difference between $\Lambda$ and $\bar{\Lambda}$ global polarization exists, the underlying differences in various theoretical calculations and spin/thermal equilibration timescale are also interesting works. Therefore, the precise measurements based on the BES-II data and FXT mode are necessary. 

Figure \ref{f10} left panel shows the energy dependence of the hyperon global polarization measurements with the newly added $\Xi$ and $\Omega$ Au+Au 200 GeV results \cite{PRLTaku}. The difference of two methods for $\Xi$ polarization extractions is within 1 $\sigma$ with given uncertainties. However, the averaged vaule of two $\Xi$ polarization extractions ($\left\langle P_{\Xi}\right\rangle(\%)=0.47 \pm 0.10 \text { (stat) } \pm 0.23 \text { (syst)}$) is larger than those for $\Lambda$ values ($\left\langle P_{\Lambda}\right\rangle(\%)=0.24 \pm 0.03 \pm 0.03$). The global polarization value of $\Omega$ was also measured to be $\left\langle P_{\Omega}\right\rangle(\%)=1.11 \pm 0.87(\text { stat }) \pm 1.97(\text { syst })$ for 20\%$-$80\% centrality. The larger hyperon polarization for more peripheral collisions indicates the increased vorticity of the system and is observed in data and are compared to the calculations from 3FD and AMPT \cite{Joey}.

\section{Future Opportunities} \label{future}
STAR has finished the scientific data taking for Run-21: 1) The highest priority is to complete the second phase of the BES-II program. 2) The second-highest priority is four short FXT runs with the detector upgrade of the iTPC and eTOF. 3) The third-highest priority is to collect data of O+O runs at $\sqrt{s_{NN}}$= 200 GeV, Au+Au runs at $\sqrt{s_{NN}}$= 17.1 GeV and 2 billion events at $\sqrt{s_{NN}}$= 3 GeV in FXT mode \cite{starData}. Thanks to the efficient RHIC operation, we would take the bonus $d$+Au runs with 100 million minimumbias and 100 million central events as shown in Table. 1. 

\begin{table}[hbpt]
\small
\tbl{STAR Run-21 efficient runs and data-taking \cite{starData}.}
{\begin{tabular}{@{}cccccc@{}} \toprule
	\hline
Single-Beam Energy & $\sqrt{s_{\mathrm{NN}}}$ & $\text {Run Time}$ &
$\text{Species}$ & $\text{Events}$ & $\text{Priority}$  \\
(GeV /\text {nucleon }) & (Rad/s) & (Rad/s) \\ \colrule
3.85\hphantom{00} & \hphantom{0}7.7 & \hphantom{0}11-20 weeks & \hphantom{0}Au+Au & \hphantom{0}100M & 1 \\
\hline
3.85\hphantom{00} & \hphantom{0}3(FXT) & \hphantom{0}3 days  & \hphantom{0}Au+Au & \hphantom{0}300M & 2 \\
44.5\hphantom{00} & \hphantom{0}9.2(FXT) & \hphantom{0} 0.5 days & \hphantom{0}Au+Au & \hphantom{0}50M & 2 \\
70\hphantom{00}   & \hphantom{0}11.5(FXT) & \hphantom{0}0.5 days & \hphantom{0}Au+Au & \hphantom{0}50M & 2 \\
100\hphantom{00}  & \hphantom{0}13.7(FXT) & \hphantom{0}0.5 days & \hphantom{0}Au+Au & \hphantom{0}50M & 2 \\
\hline
100\hphantom{00} & \hphantom{0}200 & \hphantom{0}1 week & \hphantom{0}O+O & \hphantom{0}400M + 200M(central) & 3 \\
\hline
8.35\hphantom{00} & \hphantom{0}17.1 & \hphantom{0}2.5 weeks & \hphantom{0}Au+Au & \hphantom{0}250M& 3 \\
3.85\hphantom{00} & \hphantom{0}3(FXT) & \hphantom{0}3 weeks & \hphantom{0}Au+Au & \hphantom{0}2B & 3 \\
\hline
100\hphantom{00} & \hphantom{0}200 & \hphantom{0}1 week & \hphantom{0}d+Au & \hphantom{0}100M MB + 100M(central) & 4 \\\hline
\botrule 
\end{tabular} \label{table1}}
\end{table}

In addition to critical point search, these data will enable STAR to explore, with unprecedented precision, numerous important physics. Briefly, some potential works on flow, chirality, and vorticity sides are as follows: 
\begin{itemlist}
 \item The high statistics isobar Ru+Ru and Zr+Zr collisions could be used to perform a new and compelling experimental evidence of the nuclear structure including nuclear deformation and neutron skin thickness in relativistic nuclear collisions. A more profound understanding of the $^{96}$Ru and $^{96}$Zr nuclei allows us to gain the nuclear structure and its effect on the CME search, i.e. isobar baseline and background estimations. Theoretical model calculations \cite{zhang,giuliano,jia,jia2,hanlin2,haojie1,haojie3,haojie2,Nijs,Fei,Edward,Huichao,Jun} are necessary to understand above physics. Unlike Radioactive Ion Beam Line in Lanzhou (RIBLL), Facility for Rare Isotope Beams (FRIB), and Nuclotron-based Ion Collider fAcility (NICA) at low and medium energies, isobar collisions open up new opportunity to study nuclear structure at a very short time scale ($\sim 10^{-23}s$) through heavy-ion collisions.
 \item One fundamental property of light atomic nuclei in unusual nuclear structure regimes is the $\alpha$ cluster structure \cite{JP,Wanbing}. It is a good opportunity to directly provide experimental evidence using relativistic nuclear collisions for the first time\cite{WB}. Intuitively, the configuration of $\alpha$ nucleonic cluster could be deposited in the initial state, therefore such effect could be traced via final state harmonic flow\cite{Song,Piotr2,MR,Song2,NS}. In conjunction with the measurements of nuclear deformation and neutron skin thickness, the basic understanding of the nucleon topological structure could be achieved by investigating the $\alpha$ cluster in $^{16}$O nuclei at STAR.
 \item The interpretation of a fluid-like state in small collisions has been challenged due to the small collision size and short thermalization/evolution \cite{SL}. In understanding the early-time conditions of small systems, O+O runs would allow for a direct comparison with a similarly proposed higher-energy O+O run at the LHC. Whether the small system collectivity arises from the initial momentum correlation or from the final state interaction could be distinguished \cite{STARsmall,phenix0}.
 \item There is a disagreement of triangular flow $v_3$ between STAR and PHENIX \cite{PHENIX} in the small system $p/d/^{3}$He+Au collisions. The origin of the difference has hitherto been not fully understood. More $d$+Au events with iTPC and EPD detectors could help to decipher this puzzle.
 \item To study the effect of initial state momentum correlations in small collision systems, the correlator $\rho\left(v_{2}^{2},\left[p_{T}\right]\right)$ has been proposed to be a key experimental measurement\cite{GG4,Bjoern}. The $d$+Au and O+O collision data in 2021 run provide a potential chance to prove the presence and importance of the initial state momentum anisotropies predicted by an effective theory of QCD at RHIC energies.
\item 2 billion events at Au+Au 3 GeV FXT mode providing enhanced statistics enable the measurements of proton high-order moments/cumulants. Furthermore, at lower energies, the large baryon chemical potential allows to precisely measure $\phi$ meson flow, hypernuclei lifetime, and binding energy \cite{starData}. 
 \item The large data taking in BES-II program and FXT mode is intriguing to study the polarization and vorticity of the QGP. The energy and pseudorapidity dependence of the global polarization at lower energies below 7.7 GeV would be better understood \cite{starData}.
 \item A good precision from the RHIC BES-II datasets with EPD detector providing a modern versatility for the CME search could be achieved at lower energies, where the electromagnetic field may still be larger and the flow/nonflow related background may be smaller. 
\end{itemlist}
\section{Summary}
Recently, the STAR experiment has reported important measurements: anisotropic flow in small systems, the nuclear structure probe based on flow correlations with mean transverse momentum ($\rho(v_n^2,[p_T])$) and mean transverse momentum fluctuations, the energy dependence of longitudinal decorrelations, collectivity measurements in FXT mode, CME search and vorticity/polarization measurements. Besides, based on the Run-21 efficient data-takings, future opportunities for precise measurements are also elaborated.


\section*{Acknowledgments}
Author acknowledges the STAR Collaboration, FCV PWG for the tremendous contributions, critical comments, and helpful suggestions. C. Zhang is supported by DOE Award No. DEFG0287ER40331.
\appendix


\begin{thebibliography}{0}    
\bibitem{Romatschke} R. Paul and R. Ulrike, {\it Cambridge Monographs on Mathematical Physics, Cambridge University Press,} {\bf 5} (2019).
\bibitem{Teaney} S. Schlichting and D. Teaney, {\it Annu. Rev. Nucl. Part. Sci.} {\bf 69}, 447 (2019). 
\bibitem{Heinz} U. Heinz and R. Snellings, {\it Annu. Rev. Nucl. Part. Sci.} {\bf 63}, 123 (2013). 
\bibitem{FGG} F.G. Gardim, G. Giacalone, M. Luzum and J.Y. Ollitrault, {\it Nature Phys.} {\bf 16}, 615 (2020).
\bibitem{HuichaoPRL} H.C. Song, S.A. Bass, U. Heinz, T. Hirano and C. Shen, {\it Phys. Rev. Lett.} {\bf 109}, 139904 (2012).
\bibitem{SV} S. Voloshin and Y. Zhang, {\it Z. Phys. C} {\bf 70}, 665 (1996).
\bibitem{DK} D. Kharzeev, {\it Phys. Lett. B} {\bf 633}, 260 (2006).
\bibitem{JF} D. Kharzeev, A. Krasnitz and R. Venugopalan, {\it Phys. Lett. B} {\bf545}, 298 (2002).
\bibitem{SAV} S.A. Voloshin, {\it Phys. Rev. C} {\bf 70}, 057901 (2004).

\bibitem{Liang} Z.T. Liang and X.N. Wang, {\it Phys. Rev. Lett.} {\bf 94}, 102301 (2005).
\bibitem{FB} F. Becattini and M.A. Lisa,  {\it Annu. Rev. Nucl. Part. Sci.} {\bf 70}, 395 (2020).  
\bibitem{FB2} F. Becattini and Iu. Karpenko, {\it Phys. Rev. Lett.} {\bf 120}, 012302 (2018).
\bibitem{Huichao2} B.C. Fu, S. Liu, L.G. Pang, H.C. Song and Y. Yin, {\it Phys. Rev. Lett.} {\bf 127}, 142301 (2021).
\bibitem{Nature} L. Adamczyk {\it et al.} (STAR Collaboration), {\it Nature} {\bf 548}, 62 (2017).
\bibitem{PRLTaku} J. Adam {\it et al.} (STAR Collaboration), {\it Phys. Rev. Lett.} {\bf 126}, 162301 (2021).
\bibitem{Joey} M.S. Abdallah {\it et al.} (STAR Collaboration), {\it Phys. Rev. C.} {\bf 104}, L061901 (2021).
\bibitem{Pang} L.G. Pang, H. Petersen, Q. Wang and X.N. Wang, {\it Phys. Rev. Lett.} {\bf 117}, 192301 (2016).
\bibitem{Huang} Y.C. Liu and X.G. Huang,  {\it Nucl. Sci. Tech.} {\bf 31}, 56 (2020).
\bibitem{Sun} Y.F. Sun and C.M. Ko, {\it Phys. Rev. C} {\bf 96}, 024906 (2017).
\bibitem{Deng} X.G. Deng, X.G. Huang, Y.G. Ma and S. Zhang, {\it Phys. Rev. C} {\bf 101}, 064908 (2020).

\bibitem{star0} L. Adamczyk {\it et al.} (STAR Collaboration), {\it Phys. Lett. B} {\bf 747}, 265 (2015).
\bibitem{phenix0} A. Adare {\it et al.} (PHENIX Collaboration), {\it Phys. Rev. Lett.} {\bf 111}, 212301 (2013).
\bibitem{phenix1} A. Adare {\it et al.} (PHENIX Collaboration), {\it Phys. Rev. Lett.} {\bf 115}, 142301 (2015).
\bibitem{CMS} S. Chatrchyan {\it et al.} (CMS Collaboration), {\it Phys. Lett. B} {\bf 718}, 795 (2013).
\bibitem{ALICE} S. Chatrchyan {\it et al.} (ALICE Collaboration), {\it Phys. Lett. B} {\bf 719}, 29 (2013).
\bibitem{ATLAS} G. Aad {\it et al.} (ATLAS Collaboration), {\it Phys. Rev. Lett.} {\bf 110}, 182302 (2013).
\bibitem{PHENIX} C. Aidala {\it et al.} (PHENIX Collaboration), {\it Nature. Phys.} {\bf 15}, 214 (2019).
\bibitem{STARsmall} STAR Collaboration, {\it arXiv:} 2210.11352.
\bibitem{temp} G. Aad {\it et al.} (ATLAS Collaboration),  {\it Phys. Rev. Lett.} {\bf 116}, 172301 (2016).
\bibitem{sonic} J.L. Nagle, A. Adare, S. Beckman, T. Koblesky, J.O. Koop, D. McGlinchey, P. Romatschke, J. Carlson, J.E. Lynn and M. McCumber,  {\it Phys. Rev. Lett.} {\bf 113}, 112301 (2014).
\bibitem{supersonic} P. Romatschke,  {\it arXiv:} 1502.04745.
\bibitem{ipg} B. Schenke, C. Shen and P. Tribedy,  {\it Phys. Lett. B} {\bf 803}, 135322 (2020).
\bibitem{roy} Roy A. Lacey (for the STAR Collaboration),  {\it Nucl. Phys. A} {\bf 105}, 122041 (2016).
\bibitem{Piotr} P. Bozek,  {\it Phys. Rev. C} {\bf 93}, 044908 (2016).
\bibitem{GG1} G. Giacalone, G.G. Gardim, J.N. Hostler and J.Y Ollitrault, {\it Phys. Rev. C} {\bf 103}, 024909 (2021).
\bibitem{JJ1} J.Y. Jia, S.L. Huang and C.J. Zhang,  {\it Phys. Rev. C} {\bf 105}, 014906 (2022).
\bibitem{GG5} G. Giacalone, B. Schenke and C. Shen,  {\it Phys. Rev. Lett.} {\bf 128}, 042301 (2022).
\bibitem{GG6} B. Bally, M. Bender, G. Giacalone and V. Soma,  {\it Phys. Rev. Lett.} {\bf 128}, 082301 (2022).
\bibitem{GG2} G. Giacalone, {\it Phys. Rev. Lett.} {\bf 124}, 202301 (2020).
\bibitem{GG3} G. Giacalone, J.Y. Jia and C.J. Zhang,  {\it Phys. Rev. Lett.} {\bf 127}, 242301 (2021).
\bibitem{ICNFPtalk} C.J. Zhang,(STAR Collaboration), https://indico.cern.ch/event/1025480/contributions\\/4457027/attachments/2300377/3912763/ICNFP2021\_ChunjianZhang\_STAR\_v6.pdf
\bibitem{Shou} Q.Y. Shou, Y.G. Ma, P. Sorensen, A.H. Tang, F. Videbaek and H. Wang,  {\it Phys. Lett. B} {\bf 749}, 215 (2015).
\bibitem{BS} B. Schenke, C. Shen and D. Teaney,  {\it Phys. Rev. C} {\bf 102}, 034905 (2020).
\bibitem{vnpt1} G. Aad {\it et al.} (ATLAS Collaboration), {\it Eur. Phys. J. C} {\bf 79}, 985 (2019).
\bibitem{vnpt2} S. Acharya {\it et al.} (ALICE Collaboration), {\it Phys. Lett. B} {\bf 834} 137393 (2022).
\bibitem{vnpt3} B. Somadutta (for the ATLAS Collaboration),  {\it Pos EPS-HEP2021} {\bf 398}, 0305 (2022).
\bibitem{vnpt4} C.J. Zhang, A. Behera, S. Bhatta and J.Y. Jia, {\it Phys. Lett. B} {\bf 822}, 136702 (2020).
\bibitem{vnpt5} N. Magdy an R.A. Lacey, {\it Phys. Lett. B} {\bf 821}, 136625 (2021).
\bibitem{ratiosI} C.J. Zhang, S. Bhatta and J.Y. Jia, {\it Phys. Rev. C} {\bf 106}, L031901 (2022).
\bibitem{GG4} G. Giacalone, B. Schenke and C. Shen,  {\it Phys. Rev. Lett.} {\bf 125}, 192301 (2020).
\bibitem{Bjoern} B. Schenke,  {\it Rep. Prog. Phys.} {\bf 84}, 082301 (2021).
\bibitem{SL} S.L. Huang, Z.Y. Chen, W. Li, and J.Y. Jia,  {\it Phys. Rev. C} {\bf 101}, 021901 (2020).
\bibitem{GGpt} G. Giacalone, F.G. Gardim, J.N. Hostler, J.Y. Ollitrault, {\it Phys. Rev. C} {\bf 103}, 024910 (2021).
\bibitem{JJ2} J.Y. Jia, {\it Phys. Rev. C} 105, 044905 (2022).
\bibitem{pT1} J. Adam {\it et al.} (STAR Collaboration), {\it Phys. Rev. C} {\bf 72}, 044902 (2005).
\bibitem{pT2} B. B. Abelev {\it et al.} (ALICE Collaboration), {\it Phys. Lett. B} {\bf 727}, 371 (2013).
\bibitem{pT3} B. B. Abelev {\it et al.} (ALICE Collaboration), {\it Eur. Phys. J. C} {\bf 74}, 3077 (2014).
\bibitem{pT4} L. Adamczyk {\it et al.} (STAR Collaboration), {\it Phys. Rev. C} {\bf 96}, 044904 (2017).
\bibitem{pT5} J. Adam {\it et al.} (STAR Collaboration), {\it Phys. Rev. C} {\bf 99}, 044918 (2019).
\bibitem{Som} S. Bhatta, C.J. Zhang and J.Y. Jia,  {\it Phys. Rev. C} {\bf 105}, 024904 (2022).
\bibitem{Wang} X.N. Wang and M. Gyulassy, {\it Phys. Rev. D} {\bf 44}, 3501 (1991).
\bibitem{decorr1} P. Bozek, W. Broniowski and J. Moreira,  {\it Phys. Rev. C} {\bf 83}, 034911 (2011).
\bibitem{decorr2} V. Khachatryan {\it et al.} (CMS Collaboration), {\it Phys. Rev. C} {\bf 92}, 034911 (2015).
\bibitem{decorr3} G. Aad {\it et al.} (ATLAS Collaboration), {\it Phys. Rev. Lett.} {\bf 126}, 122301 (2021).
\bibitem{decorr4} L.G. Pang, H. Petersen, G.Y. Qin, V. Roy and X.N. Wang,  {\it Eur. Phys. J. A} {\bf 52}, 97 (2016).
\bibitem{decorr5} M.W. Nie (for the STAR Collaboration), {\it Nucl. Phys. A} {\bf 1005}, 121783 (2020).
\bibitem{decorr6} J. Jia and P.Huo, {\it Phys. Rev. C} {\bf90} 034915 (2014).
\bibitem{decorr7} J. Jia, P. Huo, G. Ma and M. Nie, {\it J. Phys. G} {\bf44}, 075106 (2017).
\bibitem{bes2} M.M. Aggarwal {\it et al.} (STAR Collaboration), {\it arXiv:} 1007.2613.
\bibitem{besPRL} J. Adam {\it et al.} (STAR Collaboration), {\it Phys. Rev. Lett.} {\bf 126} 092301 (2021).
\bibitem{fxt1} K. Meehan (for the STAR Collaboration), {\it Nucl. Phys. A} {\bf 967}, 808 (2017).
\bibitem{fxt4} STAR Collaboration, {\it arXiv:} 2209.11940.
\bibitem{fxt2} J. Adam {\it et al.} (STAR Collaboration), {\it Phys. Rev. C} {\bf 103}, 034908 (2021).
\bibitem{fxt3} M.S. Abdallah {\it et al.} (STAR Collaboration), {\it Phys. Lett. B} {\bf827}, 137003 (2022).
\bibitem{DK1} D.E. Kharzeev, {\it Prog. in Part. and Nucl.} {\bf 75}, 133 (2014).
\bibitem{DK2} D.E. Kharzeev, L.D. Mclerran, H.J. Warringa, {\it Nucl. Phys. A} {\bf 803}, 227 (2008).
\bibitem{DK3} D.E. Kharzeev, J.F. Liao, S.A. Voloshin and G. Wang, {\it Prog. in Part. and Nucl.} {\bf 88}, 1 (2016).
\bibitem{CMEstar0} B.I. Abelev {\it et al.} (STAR Collaboration), {\it Phys. Rev. Lett.} {\bf 103}, 251601 (2009). 
\bibitem{CMEstar1} STAR Collaboration, {\it RHIC Beam Use Request For Runs 17 and 18}, (2016). 
\bibitem{CMEstar2} S. Choudhury {\it et al.}, {\it Chinese Phys. C} {\bf 46}, 014101 (2022).
\bibitem{CMEstar} M.S. Abdallah {\it et al.} (STAR Collaboration), {\it Phys. Rev. C} {\bf 105}, 014901 (2022). 
\bibitem{starData} STAR Collaboration, {\it The STAR Beam Use Request for Run-21, Run-22 and data taking in 2023-25}, (2020). 

\bibitem{zhang} C.J. Zhang and J.Y. Jia, {\it Phys. Rev. Lett.} {\bf 128}, 022301 (2022).
\bibitem{giuliano} G. Giacalone, J.Y. Jia and V. Soma, {\it Phys. Rev. C} {\bf 104}, L041903 (2021).
\bibitem{jia} J.Y. Jia, {\it Phys. Rev. C} {\bf 105}, 014905 (2021).
\bibitem{jia2} J.Y. Jia and C.J. Zhang, {\it arXiv:} 2111.15559.
\bibitem{hanlin2} H.L. Li, H.J. Xu,  Y. Zhou, X.B. Wang, J. Zhao, L.W. Chen and F.Q Wang, {\it Phys. Rev. Lett.} {\bf 125}, 222301 (2020).
\bibitem{haojie1} H.J. Xu, H.L. Li, X.B. Wang, C.W. Shen, F.Q. Wang, {\it Phys. Lett. B} {\bf 819}, 1136453 (2021).
\bibitem{haojie3} H.J. Xu, X.B. Wang, H.L. Li, J. Zhao, Z.W. Lin, C.W. Shen, F.Q. Wang, {\it Phys. Rev. Lett.} {\bf 121}, 022301 (2018).
\bibitem{haojie2} H.J. Xu, W.B. Zhao, H.L. Li, Y. Zhou, L.W Chen, F.Q. Wang, {\it arXiv:} 2111.14812.
\bibitem{Nijs} G. Nijs and W. Schee, {\it arXiv:} 2112.13771.
\bibitem{Fei} F. Li, Y.G. Ma, S. Zhang, G.L. Ma and Q.Y. Shou, {\it Phys. Rev. C} {\bf 106} 014906 (2022).
\bibitem{Edward} E. Shuryak, {\it arXiv:} 2201.11064.
\bibitem{Huichao} H.C. Song, {\it Flow and Spin polarization at RHIC isobar run} (RBRC workshop 2022).
\bibitem{Jun} L.M. Liu, C.J. Zhang, J. Zhou, J. Xu, J.Y. Jia and G.X. Peng, {\it Phys. Lett. B} {\bf 834} 137441 (2022).
\bibitem{JP} J.P. Ebran, E. Khan, T. Niksic and D. Vretenar, {\it Nature} {\bf 487}, 341 (2012).
\bibitem{Wanbing} W.B. He, Y.G. Ma, X.G. Cao, X.Z. Cai and G.Q. Zhang, {\it Phys. Rev. Lett.} {\bf 113}, 032506 (2014).
\bibitem{WB} W. Broniowski and E.R. Arriola, {\it Phys. Rev. Lett.} {\bf 112}, 0112501 (2014).
\bibitem{Piotr2} P. Bozek, W. Broniowski, E.R. Arrola and M. Rybczynski, {\it Phys. Rev. C} {\bf 90}, 064902 (2014).
\bibitem{Song} S. Zhang, Y.G. Ma, J.H. Chen, W.B. He and C. Zhong, {\it Phys. Rev. C} {\bf 95}, 064904 (2017).
\bibitem{MR} M. Rybczynski, M. Piotrowska and W. Broniowski, {\it Phys. Rev. C} {\bf 97}, 034912 (2018).
\bibitem{Song2} Y.A. Li, S. Zhang and Y.G. Ma, {\it Phys. Rev. C} {\bf 102}, 054907 (2020).
\bibitem{NS} N. Summerfield, B.N. Lu, C. Plumberg, D. Lee, J.N. Hostler and A. Timmins, {\it Phys. Rev. C} {\bf 104}, L041901 (2021).









\end{thebibliography}

\end{document}